\documentclass[12pt]{JHEP3}

\usepackage{amsmath,amsfonts}
\usepackage{epsfig,multicol,bbm}

\newcommand{\ud}{\mathrm{d}}

\title{Asymmetric cyclic evolution in polymerised cosmology}

\author{Orest Hrycyna$^{a}$, Jakub Mielczarek$^{b}$ and Marek
Szyd{\l}owski$^{b,c}$ \\
$^{a}$ Department of Theoretical Physics, Faculty of Philosophy,
The John Paul II Catholic University of Lublin, Al. Rac{\l}awickie 14, 20-950
Lublin, Poland \\
$^{b}$ Astronomical Observatory, Jagiellonian University, Orla 171, 30-244
Krak{\'o}w, Poland \\
$^{c}$ Mark Kac Complex Systems Research Centre, Jagiellonian University,
Reymonta 4, 30-059 Krak{\'o}w, Poland \\
E-mail: \email{hrycyna@kul.lublin.pl}, \email{jakub.mielczarek@uj.edu.pl}, 
\email{uoszydlo@cyf-kr.edu.pl}
}

\abstract{
The dynamical systems methods are used to study evolution of the polymerised 
scalar field cosmologies with the cosmological constant. We have 
found all evolutional paths admissible for all initial conditions on the 
two-dimensional phase space. We have shown that the cyclic solutions are 
generic. The exact solution for polymerised cosmology is also obtained.  
Two basic cases are investigated, the polymerised scalar field and the 
polymerised gravitational and scalar field part. In the former the division 
on the cyclic and non-cyclic behaviour is established following the sign of 
the cosmological constant. The value of the cosmological constant is upper
bounded purely from the dynamical setting.
}


\begin{document}

In the last years the procedure of background independent quantisation has been
applied to the mini-superspace cosmological models. This approach is known as
Loop Quantum Cosmology \cite{Bojowald:2008zzb}. In this theory the gravitational
part is quantised in the background independent way (\emph{polymer
quantisation}) while the matter part remains classical. The significant result
of this theory is resolution of the cosmological singularity problem. This was
strictly proved only for the models with the free scalar field. However the
considerations based on the ``effective'' equation show that it is the case also
for some models with a potential function \cite{Singh:2006im}.

Recently the procedure of polymer quantisation has been applied also to the
matter part. In their paper Hossain et al. \cite{Hossain:2009ru} constructed the
model with polymerised free scalar field. In this model the gravitational part
is treated classically. They showed that the evolution is nonsingular because the
energy density of the polymerised is always finite. Moreover the inflationary
phase emerges from this model.

The natural continuation of these studies would be construction of the fully
quantum polymerised cosmological model. In such a model both the gravity and
matter part would be quantised in the background independent way. However
Hossain et al. \cite{Hossain:2009ru} indicated the possible problems with
construction of the proper algebra for polymer variables. It is due to the fact
that matter part variables involve gravitational degrees of freedom.

In this paper we construct the classical model with the known effects of
polymerisation. The model is classical therefore the problems indicated on the
quantum level do not occur. The effects of polymerisation are introduced by the
phenomenological factors in the classical Hamiltonian. We do not state that the
model reproduces the results which could be obtained in the fully quantum
approach. Perhaps in some regimes, where quantum correlations between matter and
gravity are strong, this approximation does not work. In this approach we assume
that the gravity factors in the definitions of polymer variables of the scalar
field are the mean values. In particular in the definition of the polymerised
momentum operator we have
\begin{equation}
\hat{\pi}^{(\lambda)}_{\phi} = \frac{\langle \hat{p}^{3/2} \rangle }{\lambda} 
\sin\left(\frac{\lambda}{\langle \hat{p}^{3/2} \rangle}
\hat{\pi}_{\phi}\right)       
\end{equation}
where $\hat{p}$ is a geometric area operator. Based on this assumption we derive
``effective'' equations of motion for the system. We set the FRW symmetry of the
gravitational part and assume matter content to be a free scalar field together
with cosmological constant.
The classical phase space of this model is parametrised by the canonical variables
$(c,p,\phi,\pi_{\phi})$ which fulfil the Poisson brackets
$\left\{c,p\right\}=\frac{8\pi G \gamma}{3}$ and
$\left\{\phi,\pi_{\phi}\right\}=1$. The classical Hamiltonian for this model is
given by
\begin{equation}
\mathcal{H}= - \frac{3}{8 \pi G \gamma^2}\sqrt{p} c^2
+ \frac{ \pi_{\phi}^2}{2p^{3/2}} +p^{3/2}\frac{\Lambda}{8\pi G}.
\label{Hamiltonian} 
\end{equation}
The effect of polymerisation can be now introduced by the following replacement
in the above Hamiltonian
\begin{eqnarray}
c &\rightarrow& \frac{\sqrt{p}}{\mu}  \sin\left(\frac{\mu}{\sqrt{p}}
c \right), \\
\label{graw}
\pi_{\phi} &\rightarrow& \frac{p^{3/2}}{\lambda}
\sin\left(\frac{\lambda}{p^{3/2}} \pi_{\phi}\right).
\label{mat}
\end{eqnarray}
Based on this we obtain the phenomenological Hamiltonian $\mathcal{H}_{\text{phen}}$
which depends on the unknown parameters $\mu$ and $\lambda$. The dimension
of $[\mu]=L$ and $[\lambda]=L^2$ therefore they can be interpreted
respectively as a length and area scale of polymerisation. We do not assume here
any relation between $\mu$ and $\lambda$. The polymerisation of gravity
and matter part can have quite different energy scales. In the rest of the paper
we will investigate two models, first one, based on the polymerised
quantisation of the free scalar field, and second, where the effects of
polymerisations are included both in the gravitational sector and the free
scalar field.

We concentrate on the topological structure of the phase space and different
evolutional paths of the investigated systems which can be put in the
simple form of a dynamical system of the Newtonian type \cite{Hrycyna:2008yu}.
Within a large class of solutions
the special role is played by the cyclic solutions \cite{Xiong:2007cn,
Ashtekar:2006es} which are generic (although 
structurally unstable). They are located around a centre type critical points.

For the first model we investigate the system described by the following
Hamiltonian
\begin{equation}
\mathcal{H}_{\text{phen}} = -\frac{3}{\kappa}\frac{c^{2}}{\gamma^{2}}\sqrt{p} +
\frac{1}{2\lambda^{2}}\sin^{2}{\left(\frac{\lambda}{p^{3/2}}\pi_{\phi}\right)}p^{3/2} +
\frac{\Lambda}{\kappa}p^{3/2}
\label{ham1}
\end{equation}
where $\kappa=8\pi G$. Based on the Hamilton equation $\frac{\ud f}{\ud t} =
\left\{f,\mathcal{H}_{\text{phen}}\right\}$ together with the Hamiltonian constraint
$\mathcal{H}_{\text{phen}}\approx0$ we can derive the modified Friedman equation
\begin{equation}
H^{2} := \left(\frac{1}{2p}\frac{\ud p}{\ud t}\right)^{2} = \frac{\kappa}{3}\rho
\end{equation}
where the energy density is
\begin{equation}
\rho=
\frac{1}{2\lambda^{2}}\sin^{2}{\left(\frac{\lambda}{p^{3/2}}\pi_{\phi}\right)} +
\frac{\Lambda}{\kappa}.
\end{equation}
Here $\pi_{\phi}=\rm{const.}$ and is a parameter of the theory. Moreover the 
equation for the evolution of the field $\phi$ is given by
\begin{equation}
\frac{\ud \phi}{\ud t} =
\frac{1}{2\lambda}\sin{\left(2\frac{\lambda}{p^{3/2}}\pi_{\phi}\right)}.
\end{equation}
In order to perform dynamical analysis of the investigated system first we 
make the following change of the dynamical variable
$$
x=\frac{\lambda}{p^{3/2}}\pi_{\phi},
$$
next the time reparameterization
$$
\ud t = \sqrt{\frac{2\lambda^{2}}{3\kappa}}\frac{1}{x} \ud \tau
$$
then the modified Friedman equation can be put in the following form 
\begin{equation}
\mathcal{H}_{1} = \frac{1}{2}x'^{2} - \frac{1}{2}\sin^{2}{(x)} =
\frac{\Lambda}{\kappa}\lambda^{2}.
\label{h1}
\end{equation}
The resulting dynamical system is of the Newtonian type 
\begin{equation}
\left\{
\begin{array}{ccc}
x' & = & y,\\
y' & = & -\frac{\partial V(x)}{\partial x},
\end{array} \right.
\label{sys1}
\end{equation}
where the potential function is
\begin{equation}
V(x) = - \frac{1}{2}\sin^{2}{(x)},
\end{equation}
and the term proportional to the cosmological constant $\Lambda$ plays the role of
constant energy level.

At the finite domain of the phase space any system of the Newtonian type has the
critical points of a saddle or a centre type only. The character of a critical
point is determined from the characteristic equation of the linearization
matrix, namely
$$
l^{2}=-\frac{\partial^{2} V(x)}{\partial x^{2}}\Bigg |_{x_{0}}
$$
calculated at the critical point $x_{0}$. If $l^{2}>0$ we have a saddle type
critical point, in opposite case if $l^{2}<0$ we have a centre type critical
point.

For investigated system (\ref{sys1}) coordinates of the critical points are
$(x_{0},y_{0}) = \left(\frac{k}{2}\pi,0\right)$ , $k=0,1,2,\dots$, and their
character can be identified as: a saddle type critical points $x_{0}=k\pi$ and a
centre type critical points $x_{0}=(2k+1)\pi$ where $k=0,1,2,\dots$.

The phase portrait for this system we present in Figure~\ref{fig:1}. The
dashed trajectories correspond to the vanishing cosmological constant case.
Clearly there is a family of periodic solutions which are only possible for
the negative cosmological constant.

Equation~(\ref{h1}) can be easily integrated for different special cases.
\begin{itemize}
\item{$\Lambda=0$:
\begin{equation}
\tau + {\rm const.} = \ln{\left(\tan{\left(\frac{x}{2}\right)}\right)}
\end{equation}}
\item{$\Lambda>0$:
\begin{equation}
\tau + {\rm const.} = \frac{1}{\sqrt{2\frac{\Lambda}{\kappa}\lambda^{2}}}
F\left(x | -\frac{1}{2\frac{\Lambda}{\kappa}\lambda^{2}}\right)
\end{equation}
where $F(\cdot | \cdot)$ is an incomplete elliptic integral of the first kind.}
\item{for $\Lambda<0$ we have series of periodic solutions where the
period is given by
\begin{equation}
T_{\tau} = \frac{4}{\sqrt{1-2\frac{|\Lambda|}{\kappa}\lambda^{2}}}
F\left(\arccos{\left(\sqrt{2\frac{|\Lambda|}{\kappa}\lambda^{2}}\right)} |
\frac{1}{1-2\frac{|\Lambda|}{\kappa}\lambda^{2}}\right)
\end{equation}}
\end{itemize}

\FIGURE{
\epsfig{file=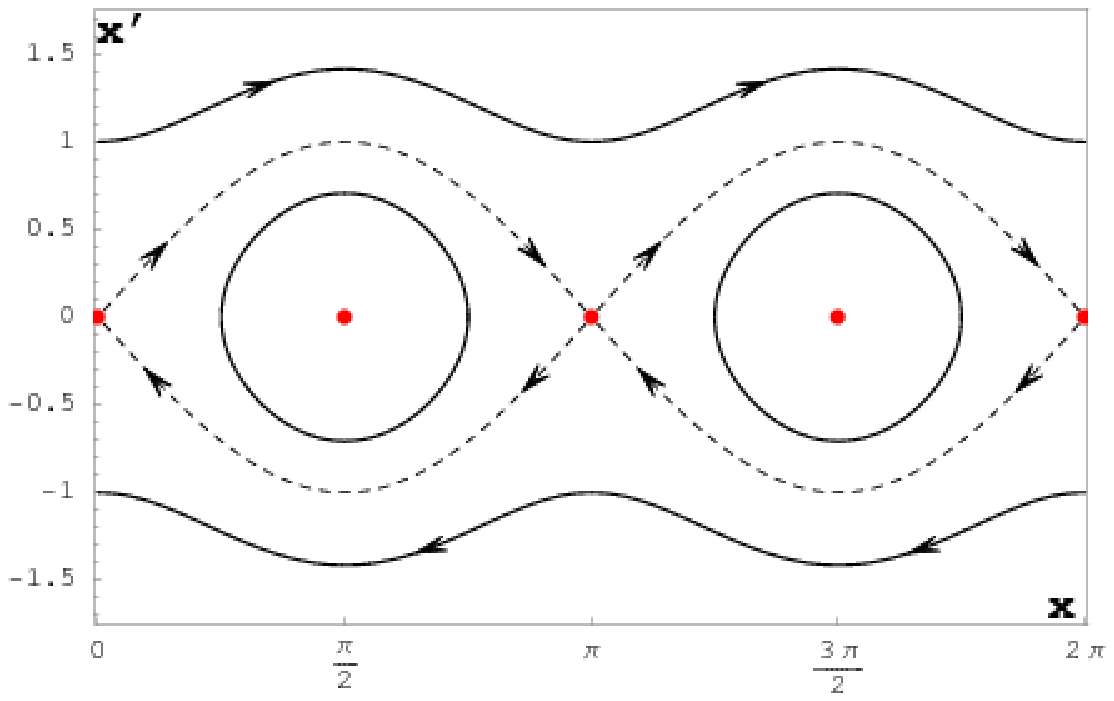,scale=0.79}
\caption{The phase portrait for the first model under consideration where the
scalar field is polymerised. All trajectories lies in the physical region. 
The dotted lines connecting saddle type critical
points denote trajectories for the vanishing cosmological constant $\Lambda=0$.
Regions around centre type critical points are occupied by the trajectories for
negative values of the cosmological constant $\Lambda<0$. The rest of the phase
space in occupied by the trajectories for $\Lambda>0$.}
\label{fig:1}
}

\FIGURE{
\epsfig{file=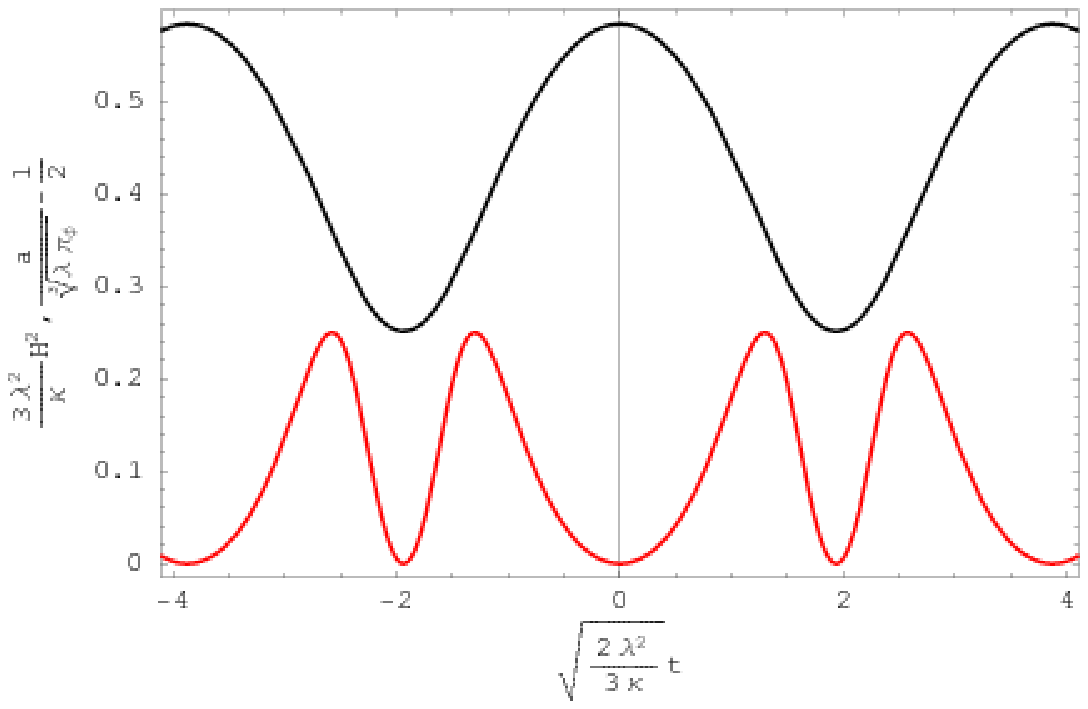,scale=0.79}
\caption{The evolution of the scale factor (black) and the Hubble function (red)
for a negative value of cosmological constant in the first model under
considerations. This is a sample of a cyclic evolution in the model. The bounce
and the recollapse take place for vanishing total energy density. This is a
specific feature of the model where the free scalar field is polymerised.}
\label{fig:4}
}

For the second model we investigate effective dynamics described by the
Hamiltonian constraint where the gravitational and free scalar field parts are
both polymerised
\begin{equation}
\mathcal{H}_{\text{phen}} =
-\frac{3}{\kappa}\frac{1}{\gamma^{2}\mu^{2}}\sin^{2}{\left(\frac{\mu}{\sqrt{p}}c\right)}p^{3/2}
+ \frac{1}{2\lambda^{2}}\sin^{2}{\left(\frac{\lambda}{p^{3/2}}\pi_{\phi}\right)}
p^{3/2} + \frac{\Lambda}{\kappa} p^{3/2}
\end{equation}

The modified Friedmann equation is in the form
\begin{equation} \label{eq:14}
H^{2}=\frac{\kappa}{3}\rho_{T}\left(1-\frac{\rho_{T}}{\rho_{c}}\right)
\end{equation}
where the total energy density
\begin{equation} \label{eq:15}
\rho_{T}=\frac{1}{2\lambda^{2}}\sin^{2}{\left(\frac{\lambda}{p^{3/2}}
\pi_{\phi}\right)} + \frac{\Lambda}{\kappa}
\end{equation}
and the critical density
\begin{equation} 
\rho_{c}=\frac{3}{\kappa\gamma^{2}\mu^{2}}.
\end{equation}

\FIGURE{
\begin{tabular}{c}
\epsfig{file=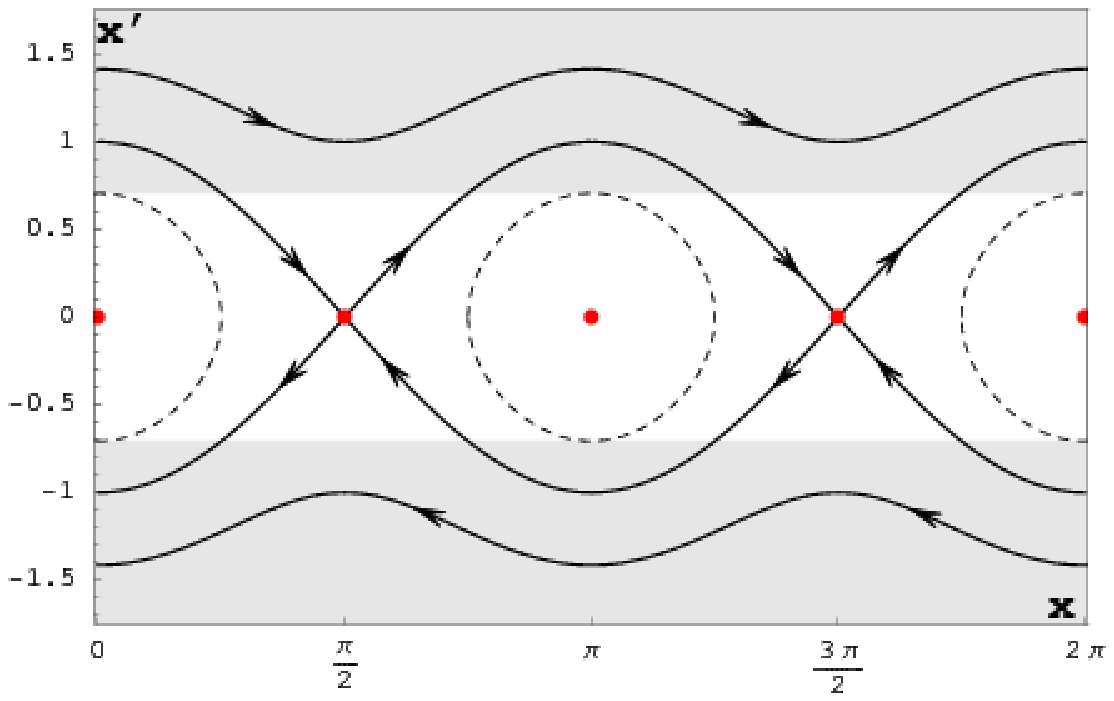,scale=0.79}\\
\epsfig{file=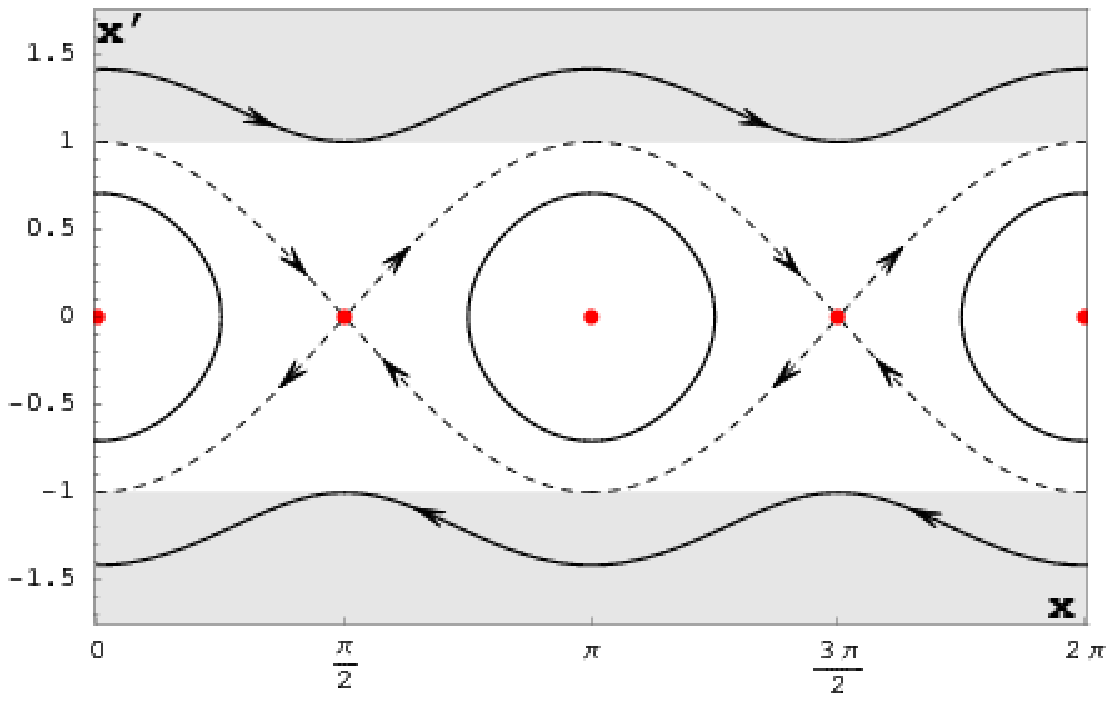,scale=0.79}\\
\epsfig{file=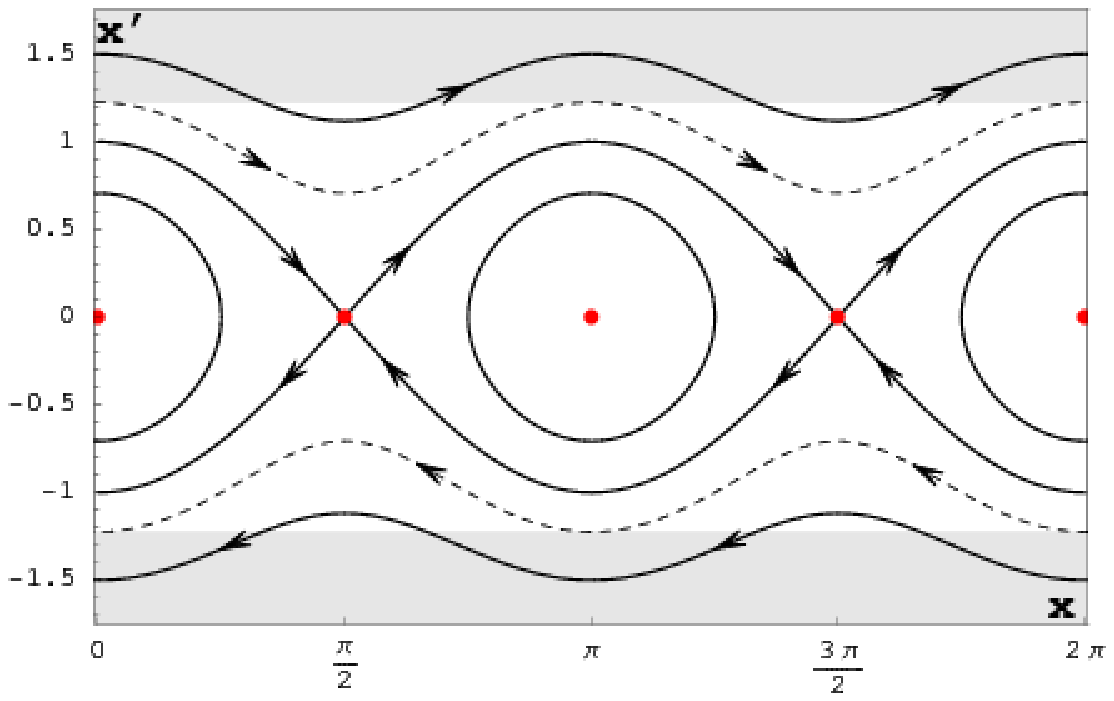,scale=0.79}
\end{tabular}
\caption{The phase diagrams for the second model under consideration, with both
gravitational and scalar field parts polymerised. From top to bottom:
$\alpha=\frac{1}{2}$, $\alpha=1$, $\alpha=\frac{3}{2}$. On the line $x=0$ $p$ is
infinite we have a singularity. The dashed lines denotes trajectories for the
vanishing cosmological constant. In the regions between them trajectories for
positive cosmological constant are located. The shaded regions are unphysical.
Every trajectory initially located in the physical domain which tend to the
unphysical region reach $H^{2}=0$ after an infinite period of the cosmological
time. Note that in all cases the trajectories for the zero cosmological constant
are tangent to the border of the unphysical region for the same value of the
coordinate $x$. The centres have coordinates $x=k\pi$, $k=1,2,\ldots$ and saddle
have coordinates $x=(2k+1)\frac{\pi}{2}$, $k=0,1,\ldots$.}
\label{fig:2}
}

Note that from (\ref{eq:14}) and (\ref{eq:15}) follows the upper limit on the
positive cosmological constant
\[
\Lambda < \kappa \rho_{\text{crit}}.
\]
Similar dynamical restriction on the value of the cosmological constant was also
found in the flat model with a scalar field and holonomy corrections of LQG
\cite{Mielczarek:2008zv,Mielczarek:2009kh}. 

The modified Friedmann equation can be put in the following form (dividing by
$\rho_{T}$)
\begin{equation}
\frac{1}{2}\frac{\dot{x}^{2}}{x^{2}}\frac{\alpha}{3\kappa\rho_{T}} =
-\frac{1}{2}\sin^{2}{(x)}+\frac{1}{2}\alpha-\frac{\Lambda}{\kappa}\lambda^{2}
\end{equation}
where $\alpha=2\lambda^{2}\rho_{c}$. After the following time reparameterization
\begin{equation}
\ud t = \sqrt{\frac{2\lambda^{2}}{3\kappa}\frac{\rho_{c}}{\rho_{T}}}\frac{1}{x}
\ud \sigma
\end{equation}
we receive
\begin{equation}
\mathcal{H}_{2} = \frac{1}{2}x'^{2} -\frac{1}{2}\left(\alpha -
\sin^{2}{(x)}\right) = -\frac{\Lambda}{\kappa}\lambda^{2}
\label{h2}
\end{equation}
In Figure~\ref{fig:2} we present the phase portraits for different values of
the parameter $\alpha$.

Note that the parameter $\alpha$ depends on both scales of polymerisation $\mu$
and $\lambda$. If we for now assume that for the gravitational part we have
standard $\bar{\mu}$ scheme then $\mu^{2}=\Delta=2 \sqrt{3} \pi \gamma l_{P}^{2}$
and
\begin{equation}
\alpha=\frac{6}{\kappa\gamma^{2}}\frac{\lambda^{2}}{\mu^{2}} =
\frac{\sqrt{3}}{8\pi^{2}\gamma^{3} l_{P}^{4}}\lambda^{2}
\end{equation}
From the phase space analysis Fig.~\ref{fig:2} we can notice that for vanishing
cosmological constant the qualitative behaviour is different for $\alpha>1$ and
$\alpha<1$. Therefore the $\alpha=1$ is a boundary value for which we can
calculate from the previous expression the boundary value of $\lambda$
parameter, namely
\begin{equation}
\lambda_{b} = \sqrt{\frac{8\pi^{2}}{\sqrt{3}}\gamma^{3}} l_{P}^{2}
\end{equation}

Next we can present investigated system (\ref{h2}) in the form of a Newtonian
type dynamical system with the potential function given by
\begin{equation}
V(x)=-\frac{1}{2}\left(\alpha-\sin^{2}{(x)}\right).
\end{equation}
As in the former case the organisation of the phase space depends only on the
form of the potential function. The location of the critical points is the same
as in the former case $x_{0}=\frac{k}{2}\pi$, $k=0,1,2,\dots$ , but their character is reversed,
namely, for $x_{0}=k\pi$ we have a centre type critical points and for
$x_{0}=(2k+1)\pi$ we have a saddle type critical points. On the Fig.~\ref{fig:2}
we present the phase space portraits for the system for different values of the
$\alpha$ parameter.

Now Eq. (\ref{h2}) can be easy integrated
\begin{equation}
\left(\frac{\ud x}{\ud \sigma}\right)^{2} =
2\left(\rho_{c}-\frac{\Lambda}{\kappa}\right)\lambda^{2} - \sin^{2}{(x)}
\end{equation}
\begin{equation}
\sigma + {\rm const.} =
\frac{1}{\sqrt{2\left(\rho_{c}-\frac{\Lambda}{\kappa}\right)\lambda^{2}}}
F\left(x|\frac{1}{2\left(\rho_{c}-\frac{\Lambda}{\kappa}\right)\lambda^{2}}\right)
\end{equation}
where $F( \cdot | \cdot)$ is an incomplete elliptic integral of the first kind.

The period of the periodic solution is given by
\begin{equation}
T_{\sigma} =
\frac{4}{\sqrt{2\left(\rho_{c}-\frac{\Lambda}{\kappa}\right)\lambda^{2}}}
F\left(\arcsin{\left(\sqrt{2\left(\rho_{c}-\frac{\Lambda}{\kappa}\right)\lambda^{2}}\right)}
| \frac{1}{2\left(\rho_{c}-\frac{\Lambda}{\kappa}\right)\lambda^{2}}\right)
\end{equation}

The unexpected feature which distinguishes this model from those meet in the
standard loop quantum cosmology is that for periodic solutions the bounce and
recollapse take place for the total energy density equal the critical density.
This happens due to presence of the periodic function $\sin^{2}{(x)}$ in the
total energy density. On the Fig.~\ref{fig:3} we present the cosmological time
evolution of three quantities proportional to the Hubble function $H^{2}$, a
difference between the total energy density and the critical density
$\rho_{c}-\rho_{T}$ and finally the scale factor $a$. Initial conditions was
chosen in such a way that at $t=0$ the universe was in its maximal expansion
phase. 

\FIGURE{
\epsfig{file=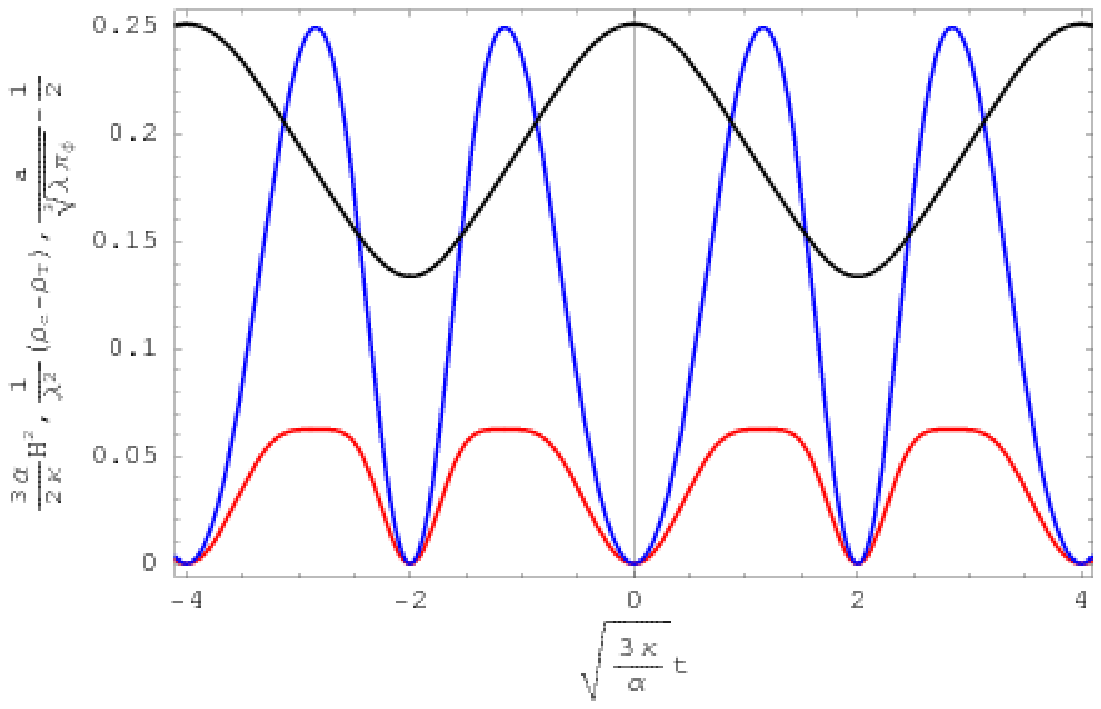,scale=0.79}
\caption{The cosmological time evolution of $H^{2}$ (red), $\rho_{c}-\rho_{T}$
(blue) and $a$ (black). The bounce and the recollapse take place for the total
energy density $\rho_{T}$ equal the critical density $\rho_{c}$.}
\label{fig:3}
}

The main conclusion of this letter is that the cyclic behaviour is generic
feature of polymerised cosmology. Two version of polymerisation has been
considered in the model with cosmological constant. First, the polymerised
scalar field only was studied, and second, both gravitation part and scalar
field have been polymerised. The cyclic trajectories are represented in the
phase space by closed phase curves around a centre type critical point. We have found
the expression for their periods as well as exact solutions for trajectories. It
is interesting that the cycle is asymmetric in original dynamical variables.
Because of the reparameterization this information is hidden.

The dynamics of the model also offers the upper limitation on the cosmological
constant value.

\providecommand{\href}[2]{#2}\begingroup\raggedright\endgroup

\end{document}